\documentclass[aps,pre,showpacs,twocolumn,floats]{revtex4}
\usepackage{amsfonts}

\usepackage{amsmath}
\usepackage{dcolumn}
\usepackage{bm}

\date{\today}

\begin{document}

\title{Acoustic decoherence of flux qubits}
\author{Jaroslav Albert and Eugene M. Chudnovsky}
\affiliation{ \mbox{Physics Department, Lehman College, City
University of New York,} \\ \mbox{250 Bedford Park Boulevard West,
Bronx, New York 10468-1589, U.S.A.}}

\begin{abstract}
Decoherence of a flux qubit due to inelastic scattering of thermal
phonons by the qubit is studied. The computed decoherence rates
contain no unknown constants and are expressed entirely in terms
of measurable parameters of the qubit. The answer depends strongly
on the size of the qubit as compared to the wavelength of phonons
of frequency $f = \Delta/(2\pi\hbar)$, with $\Delta$ being the
tunnel splitting of the qubit. Thermal phonons set the upper limit
on the operating temperature of a small qubit at around $10-20$K.
For large qubits acoustic decoherence due to one phonon processes
should be taken into account at any temperature when a high
quality factor is desired.
\end{abstract}

\maketitle

\section{Introduction.}

Flux qubits have received much attention lately due to low
decoherence exhibited at a macrosopic scale. In essence such a
qubit consists of a small superconducting loop interrupted by one
or more Josephson junctions. When half a flux quantum is applied
to the loop, the system forms a symmetric double-well potential.
The two classical states associated with the minima of this
potential correspond to clockwise and counterclockwise circulating
currents. The fascinating property of such a system is that it
exhibits quantum tunneling between the classical minima
\cite{Friedman,Mooij,Vion,Yu}, thus, providing an example of a
quantum superposition of macroscopic current states. The ground
state and the first excited state of the system are symmetric and
antisymmetric superpositions of clockwise and counterclockwise
current states, separated by the tunnel splitting $\Delta$. When
the system is prepared, e.g., in a clockwise current state, the
probability to find it with clockwise or counterclockwise current
oscillates coherently at a frequency $f = \Delta/(2 \pi \hbar)$.
Experiments have demonstrated that the quality factor of the flux
qubit can be as high as $5000$ \cite{DiVincenzo}, thus, pushing it
to the front line of promising candidates for quantum computation.
Recently, this promise has been further amplified by successful
experiments with coupled flux qubits \cite{McDermott,Majer,Hime}.

Most of the work on decoherence of flux qubits concentrates on the
effects of non-zero impedance \cite{Martinis}, electromagnetic
\cite{Friedman,Chudnovsky,Migliore,Bertet} and $1/f$ noise
\cite{Martinis,Yoshihara}. Decoherence due to coupling of a flux
qubit to the acoustics waves has been studied to a lesser degree
\cite{Chudnovsky}. Meantime, such a decoherence is generic for a
flux qubit. This can be seen from the following argument.
Tunnelling between clockwise and counterclockwise circulation is
accompanied by the reversal of the angular momentum associated
with the circular current. In experiments on flux qubits, this
angular momentum can be as large as $10^{10}\hbar$, so that its
non-conservation would be quite dramatic. A similar problem exists
for any LC circut. To conserve the angular momentum, a freely
suspended ring containing an inductor and a capacitor would have
to co-wiggle mechanically with the oscillating current. When such
a ring is attached to a solid matrix it should produce torsional
oscillations in the matrix. This immediately suggests that unless
the flux qubit is in the ground state, it should have a finite
probability to radiate a phonon. The latter can be viewed as a
consequence of the fact that the electric current is defined with
respect to the crystal lattice of ions so that quantum states of
the current are inevitably entangled with the lattice states.

It was shown earlier \cite{Chudnovsky} that the effective coupling
of the oscillating current with the crystal lattice has an
universal form that does not contain any fitting parameters. At
very low temperatures the typical decoherence rate due to this
coupling ranges from $1$s$^{-1}$ for a small qubit to
$10^6$s$^{-1}$ for a large qubit. The low temperature decoherence
rate is due to a spontaneous decay of an excited current state to
the lower energy state, accompanied by the emission of a phonon.
Such processes are known as one phonon processes. They dominate
relaxation and decoherence due to phonons in the low temperature
limit. Since operating qubits at elevated temperatures is
desirable one must consider other phonon processes which may be
dominant at these temperatures. Such processes consist of
inelastic scattering of thermal phonons by the qubit. They are
known as two phonon processes. In this paper, we focus on the
decoherence via a two phonon channel. We find that the
significance of this channel strongly depends on the size of the
qubit, $R$, as compared to the wavelength, $\lambda = v_t/f$, of a
phonon of frequency $f = \Delta/(2\pi \hbar)$ (with $v_t$ being
the speed of the transverse sound). It turns out that for a small
qubit the two phonon processes can be the main factor limiting the
qubit operating temperature. Another side of the coin, however, is
that the corresponding two phonon rate is proportional to the
square of the energy bias between the two states with opposite
directions of the circulating current. This, in principle, allows
one to switch the two phonon decoherence on and off by changing
the external magnetic flux that controls the bias. For a large
qubit the one phonon processes seem to always win over the
two-phonon processes.

The structure of the paper is as follows. Interaction between
superconducting current and elastic twists is discussed in Section
II. Two phonon processes are studied in Section III. Subsection
III-A deals with inelastic scattering of thermal phonons by small
qubits. Matrix elements and decoherence rates for large qubits are
studied in Section III-B. Section IV contains comparison of
one-phonon and two-phonon rates, as well as numerical estimates of
the acoustic decoherence rate, and general discussion of the
results.

\section{Coupling of SQUID states to acoustic phonons}

We consider a loop of radius $R$ carrying a current $J$
oscillating with frequency $f_0\sim10^9-10^{10}$s$^{-1}$ between
the clockwise and counterclockwise direction of motion. The
angular momentum associated with the circular current changes from
$L$ to $-L$ every half a period. As a consequence, the local
environment co-wiggles with the same frequency and produces
elastic distortions of the lattice in accordance with angular
momentum conservation. These distortions, ${\bf u}({\bf r}, t)$,
being pure twists, satisfy
\begin{equation}\label{Delu}
\nabla\cdot{\bf u}=0.
\end{equation}

The kinetic energy associated with the superconducting current can
be written as
\begin{equation}\label{KE}
KE=\frac{n_em_e{\bf v}_e^2}{2}=\frac{n_em_e}{2}({\bf
v}_{Lat}+\dot{\bf u})^2,
\end{equation}
where ${\bf v}_e$ is the velocity field of electrons in the
laboratory coordinate frame, ${\bf v}_{Lat}$ is same velocity in
the lattice frame, $n_e$ is the concentration of electrons, and
$m_e$ is electron mass. Interaction of the current with lattice
distortions can be formulated by noting that the current is
defined in the frame of the moving ions,
\begin{equation}
{\bf j} = en_e{\bf v}_{Lat} = en_e({\bf v}_e - \dot{\bf u}),
\end{equation}
where $e$ is electron charge. From Eq.\ (\ref{KE}) one obtains
\cite{Chudnovsky}:
\begin{equation}\label{IntHam}
{\cal H}_{int}=\frac{m_e}{e}\int d^3r{\bf j}\cdot\dot{\bf u}.
\end{equation}
This Hamiltonian satisfies all the symmetries and is free of any
unknown interaction constants.

We denote the eigenstates of the angular momentum, $L_z=L$ and
$L_z=-L$, as $|\uparrow\rangle$ and $|\downarrow\rangle$,
respectively. The system under consideration can be approximately
modeled as a particle of spin $L$ in a biased double well
potential described by the Hamiltonian ${\cal H}_0$ whose lowest
eigenstates are
\begin{equation}\label{Gstate}
|0\rangle=\frac{1}{\sqrt
2}(C_-|\uparrow\rangle+C_+|\downarrow\rangle)
\end{equation}
and
\begin{equation}\label{Estate}
|1\rangle=\frac{1}{\sqrt
2}(C_+|\uparrow\rangle-C_-|\downarrow\rangle).
\end{equation}
Here
\begin{equation}\label{CNum}
C_{\pm}=\sqrt{1\pm\varepsilon/\Delta},
\end{equation}
where
\begin{equation}
\Delta=\sqrt{\Delta_0^2+\varepsilon^2}
\end{equation}
is the energy splitting of $|0\rangle$ and $|1\rangle$,
$\varepsilon$ is the bias, and $\Delta_0$ is the energy splitting
at $\varepsilon=0$.

The total Hamiltonian of the system is
\begin{equation}\label{THam}
{\cal H}={\cal H}_0+{\cal H}_{ph}+{\cal H}_{int}.
\end{equation}
The first two terms stand for the interaction-free qubit and
phonon Hamiltonian, respectively, and the last term is given by
Eq.\ (\ref{IntHam}). Decoherence of a general superposition state
$|\Psi\rangle=c_1|0\rangle+c_2|1\rangle$ will take place through
relaxation of $|1\rangle$ to $|0\rangle$. The corresponding two
phonon transition is determined by the matrix element of ${\cal
H}_{int}$ that will be computed in the next Section. To shorten
formulas, unless stated otherwise, we will use the units in which
$k_B=\hbar=1$.

\section{Two-phonon processes}

\subsection{Small qubit.}

In the case of a small flux qubit of size $R\ll\lambda$, one can
treat the local environment at the position of the qubit as a
rigid matrix making a tiny local rotation with angular velocity
\cite{LL-elasticity}
\begin{equation}\label{AngVel}
{\bf \Omega}=\frac{1}{2}\nabla\times\dot{\bf u}.
\end{equation}
Equivalently, the velocity vector $\dot{\bf u}$ can be expressed
in terms of the angular velocity as $\dot{\bf u}={\bf
\Omega}\times{\bf r}$, where ${\bf r}$ is the position vector with
its origin at the center of the qubit. The interaction Hamiltonian
of Eq. (\ref{IntHam}) can now be simplified as
\begin{equation}\label{IntHam2}
{\cal H}_{int}={\bf L}\cdot{\bf \Omega},
\end{equation}
in which
\begin{equation}\label{L}
{\bf L}=\frac{m_e}{e}\int d^3r\ {\bf r}\times{\bf j}
\end{equation}
is the angular momentum of the circulating current. Hamiltonian of
Eq. (\ref{IntHam2}) is a consequence of the conservation of
angular momentum.

We are interested in the process of relaxation through a two phonon
channel in which one phonon is absorbed and another is emitted with
wavevectors ${\bf k}$ and ${\bf q}$ respectively. The matrix element
for this process is
\begin{eqnarray}\label{M}
M&=&\sum_{\xi}\frac{\left\langle \Psi_f\right|{\cal
H}_{int}\left|\psi_\xi \right\rangle\left\langle
\psi_\xi\right|{\cal H}_{int}\left|\Psi_i\right\rangle}{E_1\ +
\omega_{\bf k}\ - E_\xi} \nonumber \\
&+& \sum_{\xi}\frac{\left\langle \Psi_f\right|{\cal
H}_{int}\left|\psi_\xi \right\rangle\left\langle
\psi_\xi\right|{\cal H}_{int}\left|\Psi_i\right\rangle}{E_1\ -
\omega_{\bf q}\ - E_\xi} ,\
\end{eqnarray}
where $\left|\Psi_i\right\rangle$ ($\left|\Psi_f\right\rangle$)
stand for the initial (final) state of the system, $\omega_{\bf
k}$ and $\omega_{\bf q}$ are phonon frequencies, and $E_1$ is the
energy of the first excited state $|1\rangle$. The summation over
$\xi$ labels the energy states of ${\cal H}_0$, and the
intermediate phonon states $\left| n_{\bf k}-1,n_{\bf
q}\right\rangle$ in the first term and $ \left|n_{\bf k},n_{\bf
q}+1\right\rangle$ in the second term. With the help of Eq.\
(\ref{AngVel}) and conventional quantization of the phonon field,
\begin{equation}
\mathbf{u}=\sqrt{\frac{1}{2\rho V}}\sum_{\mathbf{k}\lambda }\frac{\mathbf{e}%
_{\mathbf{k}\lambda }e^{i\mathbf{k\cdot r}}}{\sqrt{\omega _{\mathbf{k}%
\lambda }}}\left( a_{\mathbf{k}\lambda }+a_{-\mathbf{k,}\lambda
}^{\dagger }\right) ,\   \label{Displacement vector}
\end{equation}
where $\rho$ is the mass density of the solid matrix and
$\mathbf{e} _{\mathbf{k}\lambda}$ is the polarization vector of
the phonon, one obtains for the phonon part of the matrix element
\begin{eqnarray}\label{PHM}
M_{ph}&=& \left\langle n_{\bf q}+1\right|\Omega_z\left|n_{\bf q}
\right\rangle\left\langle n_{\bf k}-1\right|\Omega_z \left|n_{\bf
k}\right\rangle \nonumber\\
&=&\left\langle n_{\bf k}-1\right|\Omega_z\left|n_{\bf k}
\right\rangle\left\langle
n_{\bf q}+1\right|\Omega_z\left| n_{\bf q}\right\rangle \nonumber \\
&=&\frac{1}{8\rho V}[{\bf k}\times {\bf e}_\lambda]_z[{\bf
q}\times {\bf e}_\sigma]_z\sqrt{\omega_{{\bf
k}\lambda}\omega_{{\bf q}\sigma}(n_{\bf q}+1)n_{\bf k}} . \nonumber \\
\end{eqnarray}
The action of the angular momentum operator on $|0\rangle$ and
$|1\rangle$ produces the following states:
\begin{eqnarray}\label{LS}
&&L_z|0\rangle= \frac{L\Delta_0}{\Delta}
\left[|1\rangle -\frac{\varepsilon}{\Delta_0}|0\rangle\right] \nonumber \\
&&L_z|1\rangle= \frac{L\Delta_0}{\Delta}
\left[|0\rangle+\frac{\varepsilon}{\Delta_0}|1\rangle\right].
\end{eqnarray}
Inserting these results into Eq. (\ref{M}) one can immediately see
that $\left|\psi_\xi \right\rangle$ must be either $|0\rangle$ or
$|1\rangle$. Thus, we only need to consider
\begin{eqnarray}\label{LMat}
-\left\langle 0\right|L_z\left|0\right\rangle&=&\left\langle
1\right|L_z\left|1\right\rangle=\frac{L\varepsilon}
{\Delta} \nonumber \\
\left\langle 0\right|L_z\left|1\right\rangle&=&\left\langle
1\right|L_z\left|0\right\rangle=\frac{L\Delta_0} {\Delta}.
\end{eqnarray}
Taking into account the conservation of energy, $\omega_{\bf
q}-\omega_{\bf k}=\Delta$, the matrix element (\ref{M}) reduces to
\begin{equation}\label{M2}
M=M_{ph}\frac{2L^2\varepsilon\Delta_0}{\Delta^2}\frac{1}{\omega_{\bf
k}(\omega_{\bf k}+\Delta)}.
\end{equation}
The relaxation rate can be obtained using the Fermi golden rule in
the second order, given by
\begin{equation}\label{RateDef}
\Gamma_2=\sum_{\lambda,\sigma}\int \frac{d{\bf k}d{\bf
q}}{(2\pi)^6}V^2|M|^22\pi\delta(\omega_{\bf q}-\omega_{\bf
k}-\Delta).
\end{equation}
Substitution of Eq. (\ref{M2}) into Eq. (\ref{RateDef}) yields
\begin{eqnarray}\label{Rate}
\Gamma_2&=&\frac{L^4\varepsilon^2\Delta_0^2}{16\rho^2
\Delta^2(2\pi)^5}\sum_{\lambda,\sigma}\int d{\bf k}d{\bf q}[({\bf
k}\times {\bf e}_\lambda)_z({\bf q}\times {\bf
e}_\sigma)_z]^2\nonumber \\
&&\times\frac{\omega_{\bf q}(n_{\bf q}+1)n_{\bf k}}{\omega_{\bf
k}(\omega_{\bf k}+\Delta)^2}\delta(\omega_{\bf q}-\omega_{\bf
k}-\Delta).
\end{eqnarray}
Replacing the integration variable $d{\bf k}$ with $\omega_{\bf
k}^2d\omega_{\bf k}d\Omega_{\hat{\bf k}}/v_i^2$ and the vector
${\bf k}$ with $\hat{\bf k}\omega_{\bf k}/v_i$ and $\hat{\bf
k}={\bf k}/k$, one obtains
\begin{eqnarray}\label{Rate2}
\Gamma_2&=&\frac{L^4\varepsilon^2\Delta_0^2B^2}{16\rho^2
\Delta^2(2\pi)^5}\nonumber\\
&&\times\int_0^{\omega_D} \frac{d\omega
\omega^3(\omega+\Delta)^3}{(e^\omega-1)(1-e^{-(\omega+\Delta)/T})},
\end{eqnarray}
where
\begin{equation}\label{B}
B =\sum_{\lambda}\int \frac{d\Omega_{\hat{\bf
k}}}{v_\lambda^5}(\hat{\bf k}\times{\bf e}_\lambda)_z^2.
\end{equation}
The sum in Eq. (\ref{B}) runs over the two transverse polarizations.
Making the change of
variable $x=\omega/T$ yields
\begin{equation}\label{Rate3}
\Gamma_2=\frac{L^4\varepsilon^2\Delta_0^2B^2T^7}
{16(2\pi)^5\rho^2\Delta^2}g(\Delta/T),
\end{equation}
where
\begin{equation}\label{g}
g(\Delta/T)=\int_0^{\infty} \frac{dx
x^3(x+\Delta/T)^3}{(e^x-1)(1-e^{-x-\Delta/T})}.
\end{equation}
For completeness, we insert $k_B$ and $\hbar$ into our result which
in the limit $k_BT\ll\Delta$ reduces to
\begin{equation}\label{RateFinal}
\Gamma_2\cong\frac{\pi}{270}
\frac{L^4\varepsilon^2\Delta_0^2\Delta k_B^4T^4}
{\rho^2v_t^{10}\hbar^7}.
\end{equation}
In the opposite limit $k_BT\gg\Delta$ one obtains for the rate
\begin{equation}\label{RateFinal2}
\Gamma_2\cong\frac{15\pi}{36} \frac{L^4\varepsilon^2\Delta_0^2
k_B^7T^7} {\rho^2\Delta^2v_t^{10}\hbar^7}.
\end{equation}

\subsection{Large qubit.}

For a large qubit, $R\gg\lambda$, one must integrate the
interaction of the superconducting current with phonons along the
entire loop as in Eq. (\ref{IntHam}),
\begin{equation}\label{InterHamiltLS2}
{\cal{H}}_{int}=\frac{im_e}{e}\sum_{{\bf
k}\lambda}\sqrt{\frac{\omega_{\bf k}}{2V\rho}}({\bf j}_{\bf
k}\cdot{\bf e}_i)(a_{{\bf k}\lambda}-a\dag_{{\bf k}\lambda}),
\end{equation}
where ${\bf j}_{\bf k}=\int dr^3{\bf j}e^{i{\bf k}\cdot{\bf r}}$
is the Fourier transform of ${\bf j}$. The solution to ${\bf
j}_{\bf k}$ in the thin-ring approximation is
\begin{equation}\label{jk}
{\bf j}_{\bf k}=-i2\pi RJ_1(k_\bot,R)J{\bf n}_{\bf k},
\end{equation}
where ${\bf n}_{\bf k}\perp {\bf k}$ is a unit vector in the plane
of the ring, $k_\bot$ is the z-component of ${\bf k}$, and
$J_1(k_\bot,R)$ stands for the Bessel function of the first order.
The interaction Hamiltonian can now be written as
\begin{equation}\label{InterHamiltLS3}
{\cal{H}}_{int}={\bf L}\cdot{\bf \Omega}_{eff},
\end{equation}
where
\begin{equation}
{\bf \Omega}_{eff}=-i\pi\frac{R}{a}\sum_{\bf k}
\sqrt{\frac{\omega_{\bf k}}{2V\rho}}J_1(k_\bot,R)(a_{{\bf
k}t}-a\dag_{{\bf k}t}){\bf e}_z.
\end{equation}
The scalar product ${\bf n}_{\bf k}\cdot{\bf
e}_\lambda=\delta_{t\lambda}$ as ${\bf n}_{\bf k}$ is in the plane
of the ring. To obtain the two phonon matrix element one needs
only to replace $({\bf k}\times {\bf e}_\lambda)_z({\bf q}\times
{\bf e}_\sigma)_z$ in Eq. (\ref{PHM}) with
\begin{equation}
-\left( \frac{2\pi R}{a}\right)^2J_1(k_\bot R)J_1(q_\bot R).
\end{equation}
Then, the rate for a large qubit becomes
\begin{equation}\label{Gamma1}
\Gamma_2=\frac{\pi L^4\varepsilon^2\Delta_0^2R^4}{8\rho^2 \Delta^2
a^4v_t^6} \int_0^{\omega_D}d\omega
f(\omega,\Delta,T)\Theta(\omega,R,v_t,\Delta),
\end{equation}
where
\begin{equation}
f(\omega,\Delta,T)=\frac{\omega(\omega+\Delta)}{
(e^{\omega/T}-1)(1-e^{-(\omega+\Delta)/T})},
\end{equation}
and
\begin{eqnarray}\label{Theta}
\Theta(\omega,R,v_t,\Delta)&=&\int_0^{\pi}
d\theta_1\sin\theta_1J^2_1\left(\frac{\omega
R}{v_t}\sin\theta_1\right)\nonumber\\
&\times& \int_0^{\pi}d\theta_2\sin\theta_2J^2_1\left(\frac{[\omega+\Delta] R}{v_t}
\sin\theta_2\right). \nonumber\\
\end{eqnarray}
Due to its large argument, $J_1([\omega+\Delta]R/v_t\sin\theta_2)$
can be approximated by its asymptotic form. The integral over
$\theta_2$ in Eq. (\ref{Theta}) then yields approximately $
v_t/(\omega+\Delta)R$. The integral over $\theta_1$ equals
\begin{equation}\label{P}
\frac{1}{3}\left(\frac{\omega R}{v_t}\right)^2
{_1F_2}\left(\frac{3}{2},\frac{5}{2},3,-\left(\frac{\omega
R}{v_t}\right)^2\right),
\end{equation}
where $_1F_2$ is the generalized hypergeometric function.
Inserting the results just obtained into Eq. (\ref{Gamma1}) and
making the change of variable $x=\omega/T$, one obtains
\begin{equation}\label{GammaLS}
\Gamma_2\cong \frac{\pi L^4\varepsilon^2\Delta_0^2R^5T^4}{24\rho^2
\Delta^2 a^4v_t^7}\int_0^{\omega_D/T}dx
\frac{x^3(e^x-1)^{-1}{_1F_2}(-\beta^2x^2)}{(1-e^{-x-\Delta/T})},
\end{equation}
where $\beta=RT/v_t$. In the limit $T\gg\Delta$, the integral in Eq.
(\ref{GammaLS}) simplifies to
\begin{equation}
\int_0^{\infty}dx\frac{x^3e^x{_1F_2}(-\beta^2x^2)}{(e^x-1)^2},
\end{equation}\\
or, after integrating by parts
\begin{equation}\label{XX}
\int_0^{\infty}dx \frac{x^3e^x{_1F_2}(-\beta^2x^2)}{(e^x-1)^2}=
\frac{6}{\beta^2}\int_0^{\infty}dx\frac{J_2(2\beta x)}{e^x-1}.
\end{equation}\\
For $\beta\gg1$, Eq. (\ref{XX}) reduces to $3\beta^{-2}$. The rate
then becomes
\begin{equation}\label{D}
\Gamma_2\cong \frac{\pi
L^4\varepsilon^2\Delta_0^2R^3k_B^2T^2}{8\rho^2 \Delta^2
a^4v_t^5\hbar^2},
\end{equation}
where we again inserted $k_B$ and $\hbar$ into our result. In the
limit $k_BT\ll\Delta$, one obtains
\begin{equation}\label{F}
\Gamma_2\cong \frac{\pi
L^4\varepsilon^2\Delta_0^2R^5k_B^4T^4}{24\rho^2 \Delta^2
a^4v_t^7\hbar^4}\ K\left(\frac{Rk_BT}{\hbar v_t}\right),
\end{equation}
where
\begin{equation}\label{G}
K(y)=\int_0^{\infty}\frac{x^3{_1F_2(-y^2x^2)}}{e^x-1}dx.
\end{equation}

\section{Discussion}
We are interested in the transition temperature between the rate
due to a direct process and the Raman process. The former was
calculated in Ref. \onlinecite{Chudnovsky}. It was shown that
\begin{equation}\label{DPS}
\Gamma_1=\frac{L^2\Delta^5}{12\pi \rho
v_t^5\hbar^4}\coth\left(\frac{\Delta}{2k_BT}\right)
\end{equation}
for a small qubit and
\begin{equation}\label{DPL}
\Gamma_1=\frac{ L^2 \Delta^2}{4 \pi R^3 \rho v_t^2
\hbar}\coth\left(\frac{\Delta}{2k_BT}\right)
\end{equation}
for a large qubit.

In the limit $k_BT\ll\Delta$, equations (\ref{RateFinal}) and
(\ref{DPS}) yield for a small qubit
\begin{equation}\label{Ratio1}
\frac{\Gamma_2}{\Gamma_1}=\frac{12\pi^2
L^2\varepsilon^2\Delta_0^2k_B^4T^4}{270\rho \Delta^4v_t^5\hbar^3}.
\end{equation}\\
For, e.g.,  $L\sim10^2$, $\rho\sim$5g/cm$^3$,
$v_t\sim5\times10^3$m/s, and $f_0\sim5\times10^9$s$^{-1}$, the
ratio in Eq. (\ref{Ratio1}) yields 10$^{-8}T^4$. Clearly, for
$k_BT\ll\Delta$ ($T\sim10^{-2}$K), the two phonon process is
utterly insignificant. In the opposite limit $k_BT\gg\Delta$,
equations (\ref{RateFinal2}) and (\ref{DPS}) give
\begin{equation}\label{Ratio2}
\frac{\Gamma_2}{\Gamma_1}=\frac{90\pi^2
L^2\varepsilon^2\Delta_0^2k_B^6T^6}{\rho \Delta^6v_t^5\hbar^3}.
\end{equation}
For the same parameters as considered above, Eq.\ (\ref{Ratio2})
yields $10^{-2}T^6$. The two phonon process thus dominates over a
one phonon process above $T\sim2$K. At $T \sim 20$K the
decoherence rate due to two phonon processes is of order
$10^{6}$s$^{-1}$ and, therefore, its contribution to the quality
factor of the qubit cannot be ignored.

For a large qubit in the small temperature limit one obtains with
the help of equations (\ref{F}) and (\ref{DPL})
\begin{equation}\label{Ratio3}
\frac{\Gamma_2}{\Gamma_1}=\frac{L^2\varepsilon^2\Delta_0^2k_B^4T^4}{6\pi^2\rho
\Delta^4v_t^5\hbar^3} K\left(\frac{Rk_BT}{\hbar v_t}\right).
\end{equation}
For a large qubit parameters: $L=10^{10}$, $R\sim10^{-4}$m and
$T\sim10^{-2}$K, Eq. (\ref{Ratio3}) yields $10^{-5}$. As in the
case of a small qubit, the two phonon rate is negligible compared
to the one phonon rate. In the large temperature limit equations
(\ref{D}) and (\ref{DPL}) give
\begin{equation}\label{Ratio4}
\frac{\Gamma_2}{\Gamma_1}=\frac{L^2\varepsilon^2\Delta_0^2k_BT}{4\pi^2R^2\rho
\Delta^3v_t^3\hbar},
\end{equation}
which yields $10^{-6}T$. Evidently, for a large qubit, one phonon
processes are the dominant source of decoherence at any
temperature. For the numbers used above the one phonon rate is of
order $10^{6}$s$^{-1}$.

A few important observations for flux qubits follow from the above
results. Firstly, acoustic decoherence should definitely be taken
into account when the quality factor as large as $10^4$ is
desired. Secondly, for small biased qubits the inelastic
scattering of thermal phonons by the qubit can become a dominant
mechanism of decoherence above $20$K. Finally, the proportionality
of this mechanism to the bias provides an additional way to
control the flux qubit. The interesting feature of the above
results is that they do not contain any unknown constants -- the
decoherence rate is expressed entirely in terms of measurable
parameters of the qubit.

\section{Acknowledgements}

This work has been supported by the Department of Energy through
Grant No. DE-FG02-93ER45487.

\end{document}